\documentclass[aps,prl,reprint,amsmath,showpacs,superscriptaddress]{revtex4-1}
\usepackage{graphicx}
\usepackage{textcomp}
\usepackage{braket}
\usepackage{color}
\usepackage{graphicx}
\usepackage[english]{babel}
\usepackage{SIunits}
\usepackage{color}
\usepackage{epstopdf}

\begin{document}

\title[Spin-caloric transport of optically generated spin packets in n-GaAs]{Time-resolved lateral spin-caloric transport of optically generated spin packets in n-GaAs}

\author{Stefan G\"{o}bbels}
\address{2nd Institute of Physics and JARA-FIT, RWTH Aachen University, D-52074 Aachen, Germany}
\author{Gernot G\"{u}ntherodt}
\address{2nd Institute of Physics and JARA-FIT, RWTH Aachen University, D-52074 Aachen, Germany}
\author{Bernd Beschoten}
\address{2nd Institute of Physics and JARA-FIT, RWTH Aachen University, D-52074 Aachen, Germany}

\begin{abstract}

We report on lateral spin-caloric transport (LSCT) of electron spin packets which are optically generated by ps laser pulses in the non-magnetic semiconductor n-GaAs at $T\leq35$~K. LSCT is driven by a local temperature gradient induced by an additional cw heating laser. The spatio-temporal evolution of the spin packets is probed using time-resolved Faraday rotation. We demonstrate that the local temperature-gradient induced spin diffusion is solely driven by a non-equilibrium hot spin distribution, i.e. without involvement of phonon drag effects. Additional electric field-driven spin drift experiments are used to verify directly the validity of the non-classical Einstein relation for moderately doped semiconductors at low temperatures for near band-gap excitation.

\end{abstract}


\maketitle

\section{Introduction}
Transport of spins is a crucial functional process of spintronic devices as spin injection, manipulation and detection usually take place in different parts of a device\cite{1990Datta-APL56,2004Kato-PRL93,2004Kato_Nature427,2005Kato-APL87,2005Crooker-PRL94,2007Huang-APL91,2010Wunderlich-Science330,2012_Kuhlen,2014Stepanov-APL104,Droegeler2014}. In semiconductor spintronics, the spin transport parameters (diffusivity, velocity and dephasing time) determine the length and time scale for manipulation and propagation of the non-equilibrium spin distribution after spin injection or optical excitation with excess energy above the band gap. Most important is that the spin diffusion length exceeds the device dimensions. With the discovery of the spin Seebeck effect (SSE) \cite{2008Uchida-Nature455}, a new scheme of triggering spin transport by temperature gradients was demonstrated, initiating the field of spin caloritronics\cite{2012Bauer-NMat11,2014Boona-EES7}.  The spin Seebeck physics is intensively studied in magnetic metals, semiconductors and insulators\cite{2008Uchida-Nature455,2010Jaworski-NMat9,2010Uchida-NMat9}.
The SSE can be probed in a transverse configuration where a thermal gradient is applied to a ferromagnetic material which excites the spins out of thermal equilibrium. This results in thermal spin diffusion into an adjacent normal metal perpendicular (transverse) to the direction of heat flow where it is detected\cite{Bona2014}. A thermal spin flow can also be generated across a tunneling barrier between magnetic and non-magnetic electrodes of different temperatures which results in Seebeck spin tunneling\cite{Breton2011Jun}. In this configuration, the spin flow direction is parallel (longitudinal) to the heat flow direction. A giant SSE in the non-magnetic semiconductor InSb was explained by the same processes, except that here the phonon-mediated out-of-equilibrium magnetization is carried by a magnetic field-induced spin polarization of conduction band (CB) electrons\cite{2012Jaworski_Nature487}. More recently, magnetic insulators have been used to measure the spin Peltier effect \cite{PhysRevLett.113.027601} and the spin Nernst effect \cite{NatMat16.977} in Pt/yttrium-iron-garnet heterostructures. Although other spin-caloritronic effects are also predicted for non-magnetic semiconductors\cite{2010Liu-SSC150,2010Ma-SSC150,2010Wang-SSC150}, experimental work on these materials is still lacking.

Here, we present diffusive transport of electron spin polarization in a local temperature gradient in non-magnetic n-GaAs. In contrast to the SSE observed in magnetic materials using ISHE detection, in our experimental approach, we take advantage of the spatial and temporal resolution of optical pump-probe experiments. These enable to directly map and detect lateral transport of electron spin packets driven by laser-induced local temperature gradients in n-GaAs. The basic transport feature is a spin-polarized analog of the charge-based Seebeck effect, except for the detection scheme. A fascinating and unique feature of our lateral spin-caloric transport (LSCT) under local laser heating is that the hot-electron driven spin transport is free of phonon drag contributions. The method yields direct access to the spin transport parameters, such as the spin diffusion velocity $v_\mathrm{s}$, the spin diffusion coefficient $D_\mathrm{s}$, and the spin dephasing time $T_2^*$ \cite{note1}. We show that the diffusive LSCT is corroborated by the dependence of the spin diffusion velocity $v_\mathrm{s}$ on the heating laser position, on the lattice temperature $T_\mathrm{L}$ and on the heating laser power. Further evidence stems from the independence of the LSCT-Seebeck coefficient from the lattice temperature $T_\mathrm{L}$ as well as of the spin diffusion constant $D_\mathrm{s}$ from the electric field and the overheating of the electron system $\Delta T=T_\mathrm{e}-T_\mathrm{L}$ with $T_\mathrm{e}$ being the electron temperature. The dependence of $D_\mathrm{s}$ and the spin dephasing time $T_2^*$ on the heating laser spot position gives evidence for enhanced electron densities resulting in electron-electron scattering and screening effects. The spin transport parameters are contrasted with corresponding values from our electric field-driven spin drift experiments. The latter allow to verify the non-classical Einstein relation which was predicted for moderately doped semiconductors at low temperatures, but has not been experimentally verified with respect to the electronic states involved \cite{2000Flatte-PRL84,1994Chyan-JJAP33}. The donor band (DB) states, modeled by a Gaussian density of states, play a crucial role in this validation.

\begin{figure*}[tb]
 \includegraphics{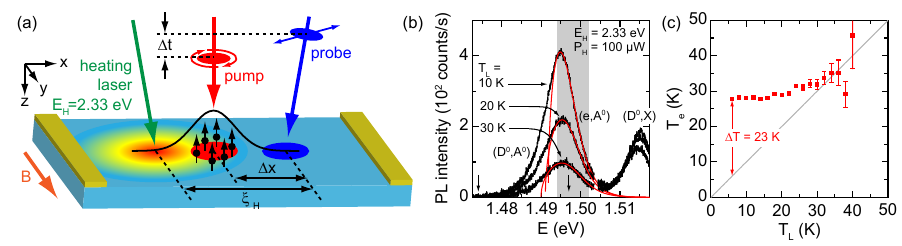}%
 \caption{\label{fig1}(a) Optical pump-probe setup to measure lateral spin transport in laser-induced temperature gradients or applied $E$ fields in GaAs. (b) PL spectra and fits of the $(e,A^0)$ transitions (red lines) at different lattice temperatures $T_\mathrm{L}$. The fit range is indicated by the grey shaded area. (c) Effective electron temperature $T_\mathrm{e}$ as a function of lattice temperature $T_\mathrm{L}$. Below $T_\mathrm{L} \approx 35$K the electron system becomes locally hotter than the lattice, creating an electron temperature gradient of approximately 0.5 K/$\mu$m at $T_\mathrm{L}$=6~K. The grey line indicates the thermal equilibrium condition $T_\mathrm{e}=T_\mathrm{L}$.}
 \end{figure*}
 
\section{Experimental setup}
\label{setup}

The GaAs sample is Si-doped close to the metal-to-insulator transition (MIT) with $n=2\times10^{16}$~cm$^{-3}$ providing spin dephasing times exceeding $100$~ns at low temperatures \cite{1998Kikkawa-PRL80,1999Kikkawa-Nature397,2002Dzhioev-PRB66,2005Crooker-PRL94,2006Furis-APL89,2007Furis-NJP9}. A specimen of $2~\mathrm{mm}\times 14~\mathrm{mm} \times 350~\upmu\mathrm{m}$ with electrical contacts at its ends is mounted in a helium flow cryostat. As depicted in Fig.1(a), electron spins are excited and oriented by circularly polarized ps pump pulses (0.5 mW) and probed by Faraday rotation $\theta_F$ of time-delayed, linearly polarized pulses (0.1 mW) from a wavelength-tunable Ti:sapphire laser\cite{PhysRevLett.105.246603}. Pump and probe pulses are focused to 24~$\upmu$m FWHM spots onto the sample with the pump being positioned at variable distances $\Delta x$ from the probe, allowing for spatio-temporal scans of the non-equilibrium spin distribution.

\subsection{Characterization of laser-induced local temperature gradients}
Local temperature gradients are created by a 100~mW cw heating laser with $E_\mathrm{H} = 2.33$\,eV that is focused to a 35\,$\upmu$m spot onto the sample at variable distances $\xi_\mathrm{H}$ from the probe spot (Fig.\ref{fig1}(a)). This yields local heating of the electron system by hot carrier excitation into the CB while leaving the lattice almost unaffected \cite{2012Kiessling-PRB86}. The local effective electron temperature $T_\mathrm{e}$ is determined from photoluminescence (PL) spectra after low intensity excitation by the heating laser with $P_\mathrm{H}$=0.1~mW, as shown in Fig.\ref{fig1}(b) for lattice temperatures $T_\mathrm{L} = 10$, 20 and 30\,K. The CB-to-acceptor transitions $(e,A^0)$ at the high-energy side of the main emission line around 1.495~eV can be fitted by spectral profiles $I(E) \propto D(E) \cdot f(E,T_e)$ (red lines), given by the CB occupation function with $D(E)$ being the density of states and $f(E,T_e)$ the Fermi function \cite{2013Quast-PRB87}. The extracted $T_\mathrm{e}$ values are plotted as a function of $T_\mathrm{L}$ in Fig.\ref{fig1}(c). The strongest increase $\Delta T = T_\mathrm{e} - T_\mathrm{L} = 23~\mathrm{K}$ is found at $T_\mathrm{L} = 6~\mathrm{K}$, decreasing for larger $T_\mathrm{L}$ and vanishing at $T_\mathrm{L}\approx 35~\mathrm{K}$ due to efficient thermalization of the hot electrons with the lattice by optical-phonon emission \cite{2012Kiessling-PRB86,1973Ulbrich-PRB8}. For the following we assume a $T_\mathrm{e}$ profile with the Gaussian shape of the heating laser spot but slightly increased width $\sigma_\mathrm{H}$ due to electronic heat diffusion, $T_\mathrm{e}(\Delta x, \xi_\mathrm{H}) = T_\mathrm{L} + \Delta T \cdot \exp{\left[-(\Delta x - \xi_\mathrm{H})^2/(2 \sigma_H^2)\right]}$ \cite{2013Quast-PRB87}. Thus, electron temperature gradients of approximately $0.5$~K/$\upmu$m can be achieved by local laser heating.

\subsection{Analysis method for LSCT}
\begin{figure*}[t]
 \includegraphics{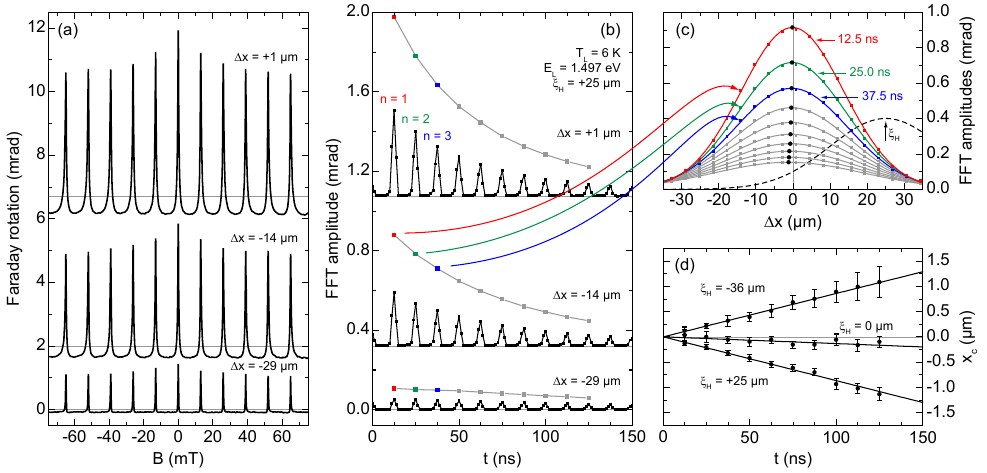}%
 \caption{\label{fig2}(a) RSA traces at $T_\mathrm{L} = 6$~K for different pump probe separations $\Delta x$ with the heating laser at $\xi_\mathrm{H} = 25~\upmu$m. (b) Corresponding FFT spectra. The amplitude of each spin packet characterized by its specific age $t_\mathrm{n}=nT_{rep}$, is given by the integrated FFT resonances (colored and grey points). Both RSA traces and FFT spectra are offset for clarity. (c) Lateral spin profiles are reconstructed from the integrated FFT resonances as indicated by the red, green and blue guide lines between panels (b) and (c). The centers of the spin packets (black dots), extracted from Gaussian fits to the lateral spin profiles (colored and grey lines), shift away from the heating laser spot (broken black line). (d) The center shifts $x_\mathrm{c}(t)$ of the spin packets change sign with the direction of the $T$ gradient which is controlled by the heating laser position $\xi_\mathrm{H}$. Lines are linear fits $x_\mathrm{c}(t) = v_\mathrm{s} t$.}
 \end{figure*}

For the investigation of LSCT in n-GaAs, a dedicated measurement and analysis method is developed allowing to observe the spatio-temporal evolution of the electron spin packets on time scales comparable to the spin dephasing time, i.e. of $T_2^* \geq$ 100~ns at low temperatures, with high spatial resolution of the spin diffusion. Due to the long spin dephasing time of $T_\mathrm{2}^*\geq 100$~ns as compared to the laser repetition interval of $T_{\mathrm{rep}}=12.5$~ns, Faraday rotation measures the superposition of many electron spin packets excited by consecutive pump pulses. The Larmor precession of the spin packets in an applied magnetic field $B$ results in resonant spin amplification (RSA) \cite {1999Kikkawa-Nature397,1998Kikkawa-PRL80,PSSB:PSSB201350201}. In Fig.\ref{fig2}(a) we show a series of RSA measurements taken at various pump-probe separations $\Delta x$ for a heating laser distance $\xi_\mathrm{H} = 25~\upmu$m (see also Fig.\ref{fig1}(a)) and at a fixed pump-probe delay of $\Delta t = -50$~ps. This yields the $B$ dependent superposition of the spin packets, $\theta_\mathrm{F}(\Delta x,B) = \sum_{n=1}^{\infty}{\theta_\mathrm{n}(\Delta x) \cdot \cos{(g \mu_\mathrm{B} B t_\mathrm{n}/\hbar)}}$, with the electron g-factor $g$, Bohr's magneton $\mu_\mathrm{B}$ and Planck's constant $\hbar$. Each spin packet is characterized by its specific age $t_\mathrm{n} = n T_{\mathrm{rep}} + \Delta t$ and precesses with its Larmor frequency $\omega = g \mu_\mathrm{B} B/\hbar$ about the $B$ field direction. Because $\left|\Delta t\right| \ll T_{\mathrm{rep}}$, the RSA traces can be treated as a Fourier series with coefficients
\begin{equation}
	\theta_\mathrm{n}(\Delta x) = \frac{\theta_\mathrm{0} \cdot e^{-t_\mathrm{n}/T_2^*} }{1+D_\mathrm{s} \, t_\mathrm{n}/\sigma^2} \cdot \exp{\left[-\frac{(\Delta x-v_\mathrm{s} \, t_\mathrm{n})^2}{4\sigma^2+4D_\mathrm{s} \, t_\mathrm{n}}\right]}
\label{eq:RSA}
\end{equation}
that describe transport of a 2D Gaussian spin packet with spin diffusion velocity $v_\mathrm{s}$, spin diffusion constant $D_\mathrm{s}$ and spin dephasing time $T_2^*$. The Gaussian width of both pump and probe laser spots is $\sigma\approx10~\upmu$m. Consequently, the fast Fourier transformation (FFT) of the RSA scans in Fig.\ref{fig2}(b) shows resonances which are the amplitudes $\theta_\mathrm{n}(\Delta x)$ of the spin packets $n = 1,~ 2,~ 3$ etc. at the pump-probe separation $\Delta x$. As each spin packet has its specific age $t_\mathrm{n}$, the corresponding FFT spectra in Fig.\ref{fig2}(b)) directly show the time evolution of the spin polarization at $\Delta x$. While the FFT amplitude decreases most rapidly at $\Delta x = 1~\upmu$m (upper curve in Fig.\ref{fig2}(b)) due to the presumably combined effects of spin dephasing and diffusion away from the center of the spin distribution at $\Delta x = 0$, the decrease becomes less at larger $\left|\Delta x\right|$ (lower curves in Fig.\ref{fig2}(b)) as the effect of spin dephasing is partially compensated by diffusion of the spins to the tail of the spin distribution. By plotting the integrated FFT amplitudes of each spin packet as a function of $\Delta x$, see Fig.\ref{fig2}(c), we can reconstruct the lateral profiles of the spin packets at times $t_\mathrm{n}$ that show their spatio-temporal evolution. The centers of the spin packets (black dots) move away from the heating laser centered at $\xi_\mathrm{H}$ (dashed black line), revealing heat-induced spin diffusion. The heating laser provides a temperature gradient as bias for the spins at $\Delta x=0$ to diffuse towards negative $\Delta x$ values. From Gaussian fits to the lateral profiles (continuous lines), the parameters $v_\mathrm{s}$, $D_\mathrm{s}$ and $T_2^*$ can be extracted from the center shifts $x_\mathrm{c}(t)=v_\mathrm{s}t$, the profile broadening $2w(t)^2=4\sigma^2+4D_\mathrm{s}t$ and the decrease of the profile amplitude $\theta_\mathrm{n}(t)=\theta_0/[1+D_\mathrm{s}t/\sigma^2] \exp{\left(-t/T_2^*\right)}$, respectively. Fig.\ref{fig2}(d) shows the temporal evolution of $x_\mathrm{c}$ of the spin packets for different heating laser positions $\xi_\mathrm{H}$. When the heating laser is moved from $\xi_\mathrm{H}=+25~\upmu$m to $\xi_\mathrm{H}=-36~\upmu$m, thereby reversing the sign of the temperature gradient $dT/dx$, this results in a reversal of the diffusion direction because the spin packets always diffuse away from the heating laser spot as expected for LSCT. At $\xi_\mathrm{H}=0$, no significant spin diffusion (center shift $x_\mathrm{c}$) is observed as the laterally averaged temperature gradient in the probe area is zero. Within the error bars, we are able to resolve center shifts of $\approx 0.5~\upmu$m even with laser spots of Gaussians widths of $\sigma \approx 10~\upmu$m.
Systematic errors of the spin transport parameters due to idealizations in the analysis model, such as the neglect of the finite pump-probe delay $\Delta t =-50$~ps and of the magnetic field dependence of the spin dephasing time $T_2^*$, which can be seen by the decrease of the RSA resonances with the magnetic field in Fig.\ref{fig2}(a), are quantified by spin transport simulations presented in the Supplements, together with further details of the FFT analysis. These systematic errors are less than $7 \%$ of the input parameters of the simulations, confirming the good reliability of our analysis method.

\section{Experimental results}
\subsection{Hot-electron driven lateral spin-caloric transport (LSCT)}

\begin{figure}[t]
 \includegraphics[width=\columnwidth]{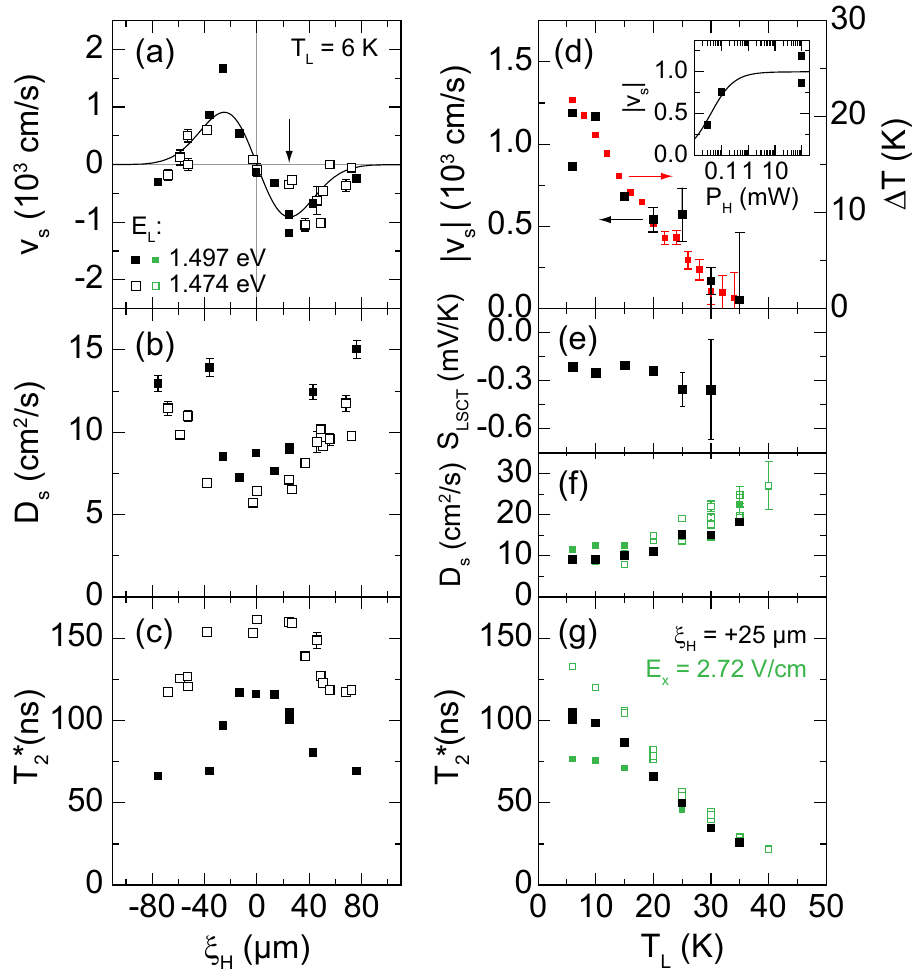}%
 \caption{\label{fig3} (color online). (a)-(c) Spin transport parameters $v_\mathrm{s}$, $D_\mathrm{s}$ and $T_2^*$ as a function of the heating laser position $\xi_\mathrm{H}$ for spin excitation energies $E_{\mathrm{Laser}} = 1.497$~eV (filled squares) and 1.474~eV (open squares) at $T_\mathrm{L} = 6$~K. The black line in (a) is a calculation by Eq.(\ref{eq:dTdx}). (d)-(g) $\left|v_\mathrm{s}\right|$, $S_{\mathrm{LSCT}}$, $D_\mathrm{s}$,  and $T_2^*$ extracted from LSCT measurements with $\xi_\mathrm{H} = 25~\upmu$m (black squares) and from electric field-driven spin transport measurements with $E_\mathrm{x} = 2.72$~V/cm (green squares) as a function of the lattice temperature $T_\mathrm{L}$. The red data points in (d) are the effective electron temperature increase $\Delta T$ measured from PL, which exhibits the same $T_\mathrm{L}$ dependence as $\left|v_\mathrm{s}\right|$. Inset: dependence of $\left|v_\mathrm{s}\right|$ on the heating laser power $P_\mathrm{H}$. The black line is a plot of Eq.(\ref{eq:powerdependence}).}
 \end{figure}

We now explore the dependence of the hot-electron driven LSCT on the temperature gradient by investigating the behavior of the spin transport parameters $v_\mathrm{s}$, $D_\mathrm{s}$ and $T_\mathrm{2}^*$ on the heating laser position $\xi_\mathrm{H}$ in Fig.\ref{fig3}(a)-(c). LSCT is measured for spin excitation both into the CB states with pump energy $E_{\mathrm{L}}=1.497$~eV (filled squares) and into the donor band (DB) states with $E_{\mathrm{L}}=1.474$~eV (open squares), corresponding to the $(e,A^0)$ and $(D^0,A^0)$ transitions in Fig.\ref{fig1}(b), respectively. For both spin excitation energies, $\left|v_\mathrm{s}\right|$ first increases with increasing distance $\xi_\mathrm{H}$ of the heating laser, reaching a maximum of $\left|v_\mathrm{s}\right| \approx 1000$~cm/s at $\left|\xi_\mathrm{H}\right| \approx 25~\upmu$m, and decreases for larger $\left|\xi_\mathrm{H}\right|$. The sign of $v_\mathrm{s}$ changes with the inversion of $\xi_\mathrm{H}$. This behavior results from the profile of the local temperature gradient which is averaged over the probe spot at $\xi_\mathrm{H} = 0$. The black line in Fig.\ref{fig3}(a) is a plot of
\begin {equation}
	v_\mathrm{s} \propto \left<-dT/dx\right> \propto \Delta T \cdot \left(-\xi_\mathrm{H}\right) \exp{\left[-\frac{\xi_\mathrm{H}^2}{2(\sigma_\mathrm{H}^2+\sigma^2)}\right]}
\label{eq:dTdx}
\end{equation}
with $\sigma_\mathrm{H} = 23~\upmu$m (corresponding to $54~\upmu$m FWHM) and $\sigma = 9~\upmu$m being the width of the pump beam. The data clearly follow this model which confirms the observation of LSCT. In contrast to $v_\mathrm{s}$, both $D_\mathrm{s}$ and $T_2^*$ show a symmetric dependence on $\xi_\mathrm{H}$. While $D_\mathrm{s}$ is lowest around $\xi_\mathrm{H} = 0$ and increases with larger distance of the heating laser, the opposite behavior is observed for $T_2^*$.  This is surprising, as $T_2^*$ is known to decrease with higher T \cite{1998Kikkawa-PRL80} which is also seen in our sample (Fig.\ref{fig3}(g)). Within the Dyakonov-Perel (DP) spin dephasing mechanism \cite{DP1971}, such an enhancement of $T_2^*$ at the heating area might result from enhanced electron-electron scattering which reduces the momentum scattering time between scattering events and thus increases the spin dephasing time. DP spin dephasing also predicts that the spin dephasing time should increase with increasing magnetic field. However, such dependency has never been observed in n-GaAs for doping concentrations close to the metal-to-insulator transition \cite{1998Kikkawa-PRL80}. The overall larger $D_\mathrm{s}$ values and lower $T_2^*$ times observed for $E_{\mathrm{L}} = 1.497$~eV (filled squares in Fig. 3(b,c)) indicate that the spins might occupy CB states, while for 1.474~eV (open squares) less mobile states in the donor band (DB) might be excited \cite{2002Dzhioev-PRB66,2010Roemer-PRB81,2012Intronati-PRL108}.

The strongest verification of hot-electron driven LSCT is given by the dependence of $v_\mathrm{s}$ on the lattice temperature $T_\mathrm{L}$ (black squares in Fig.\ref{fig3}(d)) which was measured with the heating laser at $\xi_\mathrm{H} = +25~\upmu$m, i.e. at maximum $v_\mathrm{s}$ (see arrow in Fig.\ref{fig3}(a)). Most strikingly, $\left|v_\mathrm{s}\right|$ (black squares) shows the same dependence on $T_\mathrm{L}$ as the effective electron temperature increase $\Delta T$ deduced from PL measurements (red data points), confirming eq.(\ref{eq:dTdx}). Although PL was measured under low intensity ($P_\mathrm{H} = 0.1$~mW) of the heating laser (compared to $P_\mathrm{H} =100$~mW used for the LSCT measurements), the inset of Fig.\ref{fig3}(d) indicates that spin diffusion is already in saturation for $P_\mathrm{H}\geq 0.1$~mW, justifying the comparison of $\left|v_\mathrm{s}\right|$ and $\Delta T$. The saturation results from the fact that the mean electron excess energy per absorbed photon from the heating laser, $\varepsilon$, will be rapidly distributed among the whole electron system due to electron-electron scattering which leads to both an increase of the effective electron temperature and the effective chemical potential. When these electrons diffuse away from their excitation spot they are out of equilibrium with the lattice \cite{2012Kiessling-PRB86}. Therefore, the laser-induced increase of the effective electron temperature is given by
\begin{equation}
	\Delta T \propto \frac{n_\mathrm{H} \varepsilon}{n+n_\mathrm{H}} \propto \frac{P_\mathrm{H}}{P_\mathrm{S}+P_\mathrm{H}}
\label{eq:powerdependence}
\end{equation}
with $n$ and $n_\mathrm{H}$ denoting the electron densities from doping and from optical excitation, respectively, and $P_\mathrm{S}$ being the saturation power. The black line in the inset of Fig.\ref{fig3}(d) is a plot of eq.(\ref{eq:powerdependence}) with $P_S = 40~\upmu$W. The dependencies of $v_\mathrm{s}$ on the heating laser position $\xi_\mathrm{H}$, on the lattice temperature $T_\mathrm{L}$ and on the heating laser power $P_\mathrm{H}$ prove that spin diffusion originates from a local temperature gradient generated by a hot electron distribution as described by eqs.(\ref{eq:dTdx}) and (\ref{eq:powerdependence}).

\begin{figure}
 \includegraphics[width=\columnwidth]{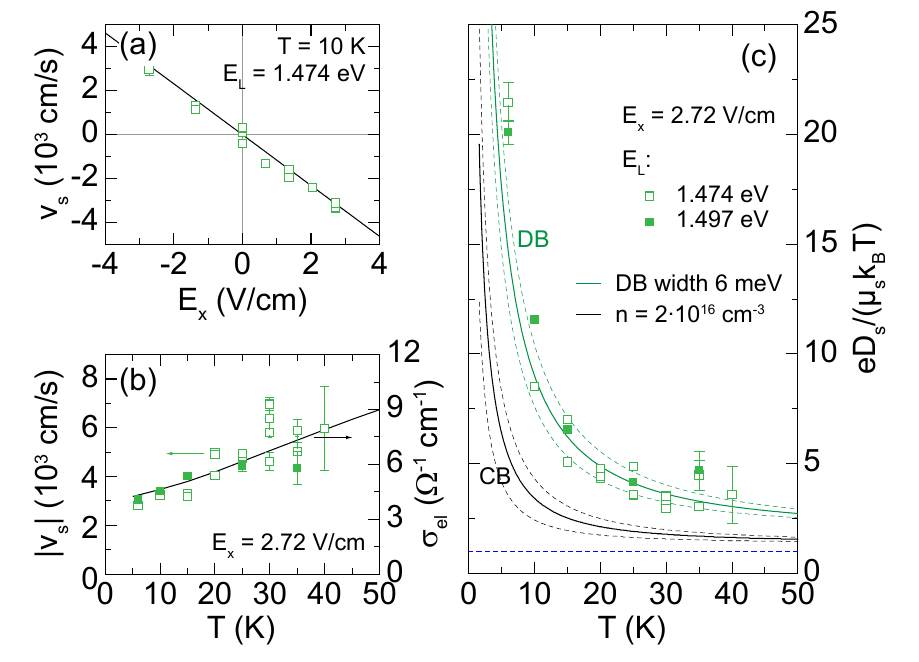}%
 \caption{\label{fig4} (color online). (a) Linear dependence of electric field-driven spin drift velocity $v_\mathrm{s}$ on $E_\mathrm{x}$ with spin mobility $\mu_\mathrm{s} = 1160$~cm$^2$~V$^{-1}$~s$^{-1}$. (b) $\left|v_\mathrm{s}\right|$ shows the same temperature dependence as the electrical conductivity $\sigma_{\mathrm{el}}$ (black line). (c) Measured and simulated Einstein relation $\mathcal{R} = eD_s/(\mu_\mathrm{s} k_\mathrm{B} T)$. The blue broken line is the classical case $\mathcal{R} = 1$, while the black and green lines are simulations of the generalized Einstein relation for CB and DB transport, respectively.}
 \end{figure}

The Seebeck coefficient of the hot-electron driven LSCT, $S_{\mathrm{LSCT}}$, can be estimated by comparing thermally-driven spin diffusion velocities with those from electric field-driven spin drift measurements. The latter were additionally carried out with electric fields $E_\mathrm{x}$ applied along the $x$-direction without the heating laser (see Fig.\ref{fig1}(a)). Fig.\ref{fig4}(a) shows a linear dependence of $v_\mathrm{s} = -\mu_\mathrm{s} E_\mathrm{x}$ with a spin mobility of $\mu_\mathrm{s} = 1160$~cm$^2$V$^{-1}$s$^{-1}$. The maximum spin diffusion velocity $\left|v_\mathrm{s}\right| \approx 1000$~cm/s of the LSCT measurements in Fig.\ref{fig3}(a), which was measured at $\xi_\mathrm{H} = 25~\upmu$m with $\left<dT/dx\right> \approx 0.5$~K/$\upmu$m, is obtained in our electric field-driven experiments at $E_\mathrm{x} = 0.86$~V/cm (see Fig.~\ref{fig4}(a)), from which we estimate $\left|S_{\mathrm{LSCT}}\right| \approx 170~\upmu$V/K at $T_\mathrm{L}=6$~K. These values are well below typical Seebeck coefficients $|S|$ in GaAs exceeding 500$\upmu$V/K at 10~K \cite{WarrenAPL}. The small dependence of $\left|S_{\mathrm{LSCT}}\right|$ on the lattice temperature $T_\mathrm{L}$ in Fig.\ref{fig3}(e) reaffirms that our LSCT is purely driven by a hot electron distribution without involvement of phonon drag effects.  This is a unique feature of our experimental approach using local laser heating, in contrast to typical charge-based Seebeck effect measurements where phonon-drag results in a distinct extremum in the temperature dependent Seebeck coefficient at low temperature which we do not observe in our LSCT measurements\cite{WarrenAPL,2008Homm-APL93}.

The dependence of $D_\mathrm{s}$ on $T_\mathrm{L}$ (Fig.\ref{fig3}(f)) is the same for both LSCT (black squares) and electric field-driven experiments (open and filled green squares, see inset of Fig. 3 (a)) indicating that spin diffusion is independent of $\Delta T$ and $E_\mathrm{x}$. The measured $T_2^*$ times in the electric field-driven spin transport  (Fig.\ref{fig3}(g), green squares) differ increasingly from the LSCT data (black squares) for both spin excitation energies starting from $T_\mathrm{L} = 25$~K towards $T_\mathrm{L} = 6$~K. At $T_\mathrm{L} = 6$~K this deviation is in agreement with the thermal equilibrium values of $T_2^*$ for both excitation energies at large $\left|\xi_\mathrm{H}\right|$ in the LSCT measurements in Fig.\ref{fig3}(c). This indicates that different spin states are excited and probed for $E_{\mathrm{L}} = 1.497$~eV and 1.474~eV which become indistinguishable at elevated temperatures ($T_\mathrm{L} \geq 25$~K), independent of the excitation energy and the heating laser, most likely resulting from thermal energy redistribution.

\subsection{Testing the Einstein relation using spin drift transport parameters. }

A closer insight into which states are involved in the measured spin transport can be gained from the temperature dependence of electric field-induced spin transport. We first note that $\left|v_\mathrm{s}\right|$ shows the same increase with $T$ (here $T_\mathrm{e} = T_\mathrm{L}$) in Fig.\ref{fig4}(b) as the electrical conductivity $\sigma_{\mathrm{el}}$ (black line) measured by 2-point $I$-$V$-curves. This implies that the electron density does not dependent on temperature as expected for itinerant CB or DB charge carriers. From the spin drift velocities in Fig.\ref{fig4}(b) together with the spin diffusion constant in Fig.\ref{fig3}(f) we can test the Einstein relation $\mathcal{R} \equiv eD_s/(\mu_\mathrm{s} k_\mathrm{B} T)$ that was predicted to deviate from the classical value $\mathcal{R} = 1$ for non-degenerate particles, i.e. doped semiconductors close to the MIT at low temperatures  \cite{2000Flatte-PRL84,1994Chyan-JJAP33}. The results in Fig.\ref{fig4}(c) verify the predicted increase of $\mathcal{R} \gg 1$ at low temperatures for fermionic particles. The blue broken line shows the classical case for Boltzmann-distributed systems. The the black solid and broken lines are simulations of the generalized Einstein relation according to Ref. \citenum{2000Flatte-PRL84} for CB transport of Fermi-Dirac-distributed electrons with the Fermi-energy $E_\mathrm{F} = \hbar^2/(2m_{\mathrm{CB}}^*) (3\pi^2n)^{2/3}$ for carrier densities $n$ between $1\times10^{16}$cm$^{-3}$ (lower curve) and $3\times10^{16}$cm$^{-3}$ (upper curve). Remarkably, all experimental data (green squares) clearly exhibit consistently larger $\mathcal{R}$ values and thus do not follow simple CB transport. In contrast, our data can be described much better by DB transport over the whole temperature range when assuming a Gaussian density of states of width $\Delta E = 6$~meV with $E_\mathrm{F}$ in the center of the band (see green solid line in Fig.\ref{fig4}(c)). For comparison, we included respective curves with $\Delta E = 5$~meV (lower green broken line) and 7~meV (upper green broken line). Surprisingly, even spins which are directly excited into the CB at $E_L=1.497$~eV (full green squares in Fig.\ref{fig4}(c)) are best described by this DB approach despite their slightly different spin dephasing times at low temperatures as discussed for Fig.\ref{fig3}(g). Alternative interpretations like ambipolar diffusion  \cite{2000Flatte-PRL84} or spin Coulomb drag \cite{PSSB200945338} have been ruled out \cite{Quast2017}.

\section{Conclusions}

In conclusion, we have demonstrated lateral spin-caloric transport of optically generated spin packets in n-type non-magnetic GaAs at low temperatures. This diffusive spin transport is driven by local temperature gradients from a hot electron distribution due to laser heating. By mapping the spatio-temporal evolution of the spin packets, our experiments give access to the spin transport parameters and thus a microscopic insight into the spin analog of the charge-based transport in the Seebeck effect. As a unique feature of our experimental approach, the driving mechanism of the LSCT is solely governed by excitation of a local hot electron distribution in the CB and therefore free of significant phonon-drag contributions yielding a nearly temperature-independent Seebeck coefficient of the LSCT. Furthermore, we tested the non-classical Einstein relation for moderately doped semiconductors and found a temperature dependence which can be well described by donor band transport. Our results strongly suggest that also thermally-induced transverse spin currents due to spin Nernst effect might be observable by our all-optical spin transport measurement technique.


Work supported by the DFG through SPP 1538.

\section*{References}

%



\widetext
\pagebreak
\begin{center}
	\textbf{\large Supplemental Material: Time-resolved lateral spin-caloric transport of optically generated spin packets in
n-GaAs}
\end{center}
\setcounter{equation}{0}
\setcounter{figure}{0}
\setcounter{table}{0}
\setcounter{page}{1}
\makeatletter
\renewcommand{\theequation}{S\arabic{equation}}
\renewcommand{\thefigure}{S\arabic{figure}}
\renewcommand{\thesection}{S\arabic{section}}
\renewcommand{\thetable}{S\arabic{table}}

\subsection*{1. Spin transport simulations: Introduction}

In the Supplement, we quantify systematic errors on the spin transport parameters with the help of
spin transport simulations. The systematic errors arise 1.) from the method of the fast Fourier transformation (FFT) that is used to reconstruct the spatio-temporal evolution of the spin packets from spatially-resolved resonant spin amplification (RSA) measurements and 2.) from idealizations in the analysis model that is used to determine the spin transport parameters.

The spatially-resolved RSA signal resulting from the superposition of Gaussian spin packets from adjacent pump pulses, including spin dephasing, spin precession as well as 2D spin drift and diffusion, is described by
\begin{equation}
	\theta_F(\Delta x,\Delta t, B) = \sum_{n=1}^{\infty}{\frac{\theta_0}{1+D_s t_n / \sigma^2} \cdot e^{-t_n / T_2^*} \cdot \exp{\left[-\frac{(\Delta x-v_s t_n)^2}{4\sigma^2+4D_s t_n}\right]} \cdot \cos{\left(g \frac{\mu_B}{\hbar} B t_n\right)}}
\label{eq:RSAfull}
\end{equation}
%
with $t_n = n \cdot T_{\mathrm{rep}} + \Delta t$. For the special case of $\Delta t = 0$, the RSA signal forms an ideal Fourier series,
\begin{eqnarray}
	&& \theta_F(\Delta x, B) = \sum_{n=1}^{\infty}{\theta_n\left(\Delta x\right) \cdot \cos{\left(n \omega_B B\right)}}\\
	\textnormal{with}\ \ && \theta_n\left(\Delta x\right) = \frac{\theta_0}{1+D_s n T_{\mathrm{rep}}/ \sigma^2} \cdot e^{-n T_{\mathrm{rep}} / T_2^*} \cdot \exp{\left[-\frac{(\Delta x-v_s n T_{\mathrm{rep}})^2}{4\sigma^2+4D_s n T_{\mathrm{rep}}}\right]}\\
	\textnormal{and} \ \ && \omega_B = g \frac{\mu_B}{\hbar} T_{\mathrm{rep}}.
\label{eq:RSAFourierForm}
\end{eqnarray}
Consequently, by reconstruction of the lateral profiles $\theta_n(\Delta x)$ of the spin packets via Fourier transformation of the RSA scans at the various pump-probe separations $\Delta x$ (as described in the main text), we can extract the spin drift velocity $v_s$, the spin diffusion constant $D_s$ and the spin dephasing time $T_2^*$ from fits to the center shifts $x_c(t_n) = v_s t_n$, to the broadening $2w(t_n)^2 = 4\sigma^2 + 4 D_s t_n$ and to the decrease of the amplitudes $\theta(t_n) = \theta_0/[1+D_s t_n/\sigma^2] \exp{\left(-t_n/T_2^*\right)}$, respectively.

Yet, this special case $\Delta t = 0$ cannot be realized experimentally, for two reasons: On the one hand, the zero order spin packet excited at time $t_0 = 0$ would be detected by the probe spot, while the sum in a Fourier series runs from $n = 1$. On the other hand, the Faraday signal around $\Delta t = 0$ is usually disturbed by some interference effects as pump and probe spots are overlapping spatially and temporally. As a result, a non-zero pump-probe delay (typically $\Delta t = -50$~ps) has to be used for the RSA measurements, such that the experimental RSA traces only approximate a Fourier series because of $\left|\Delta t\right| \ll T_{\mathrm{rep}}$. Furthermore, as the coefficients of a Fourier series must not depend on the variable, which is in our case the magnetic field $B$, another unavoidable deviation of the measured RSA traces from an ideal Fourier series is due to the magnetic field dependence of the spin dephasing time $T_2^*(B)$, which can be seen in the decrease of the RSA resonance height with increasing field strength $\left|B\right|$.

In the following section, we quantify the influence of a finite pump-probe delay $\Delta t = - 50$~ps and a magnetic field-dependent spin dephasing time $T_2^*(B)$ on the spin transport parameters resulting from our analysis model, which is based on the FFT of ideal Fourier series (i.e. which assumes $\Delta t = 0$ and magnetic field-independent spin transport parameters). For this purpose, our analysis model is applied to simulations of RSA traces for realistic measurement conditions according to eq. (\ref{eq:RSAfull}), with $\Delta t \leq 0$ and a phenomenological field-dependence $T_2^*(B)$, and the resulting parameters are compared to the input parameters. These spin transport simulations, presented below, allow us to find the optimal FFT conditions that give the best reproduction of the input parameters, i.e. the most realistic values of $T_2^*$, $v_s$ and $D_s$.

\subsection*{2. Spin transport simulations for finite pump-probe delays $\Delta t \leq 0$}
\begin{figure}[tb]
 \includegraphics{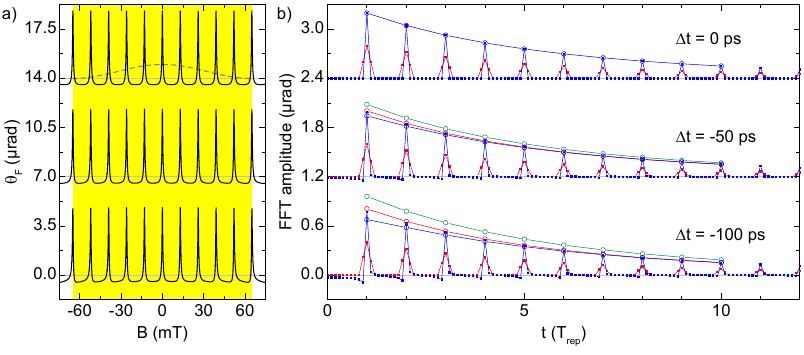}%
 \caption{\label{figSupp1} (a) Simulated RSA traces at $\Delta x = 0$ for pump-probe delays $\Delta t = 0$, -50~ps and -100~ps (upper, middle and lower curve, respectively). The yellow area shows the FFT range, including 10 RSA periods. The broken grey line indicates the weighting of the data if an optional Hann window is used for the FFT routine. (b) FFT spectra of the simulated RSA traces. Blue squares are the real parts and open blue circles the corresponding integrated resonance values of the FFT spectra without Hann window. Green and red circles are the integrated resonance values of the FFT power spectra without and with Hann window, respectively.}
\end{figure}

The simulations of spatially-resolved RSA traces were done with LabView, numerically calculating the sum of 100 spin packets according to eq. (\ref{eq:RSAfull}) for each magnetic field value in the range $-80$~mT $\leq B \leq +80$~mT and for each pump-probe separation $-31~\upmu$m $\leq \Delta x \leq +31~\upmu$m. The input parameters were chosen to be
\begin{eqnarray}
	&&\theta_0 = 1, \ \ \sigma = 10~\upmu\textnormal{m}, \ \ T_{\mathrm{rep}} = 12.5~\textnormal{ns}, \ \ g = -0.44, \ \ T_2^* = 120~\textnormal{ns},\ \ v_s = 3480~\textnormal{cm}/\textnormal{s} \ \ \nonumber \\ &&\textnormal{and} \ \ D_s = 10~\textnormal{cm}^2/\textnormal{s}
\label{eq:InputParameters}
\end{eqnarray}
close to the values measured for electrically induced spin transport at $T = 10$~K. Simulations were performed with pump-probe delays ranging from $\Delta t = 0$ (ideal case) to $\Delta t = -100$~ps. Simulated RSA traces at pump-probe-overlap $\Delta x = 0$ are shown in Fig. \ref{figSupp1}~(a) for $\Delta t = 0$, -50~ps and -100~ps (upper, middle and lower curve, respectively). While for the ideal case $\Delta t = 0$ all RSA resonances are symmetric with respect to the resonance fields $B_{\mathrm{res},m} = m \cdot 2\pi\hbar/(g \mu_B T_{\mathrm{rep}}) = m \cdot 12.99$~mT ($m$ being an integer number), the RSA resonances for non-zero pump-probe delays $\Delta t< 0$ become more and more asymmetric at higher fields.

The FFT analysis of both the measurements presented in the main text and the spin transport simulations described here was done with a standard LabView FFT routine. Prior to the FFT, each RSA trace was first interpolated to equidistant $B$-steps (necessary for the FFT routine) of $\Delta B = 0.01$~mT. Then the $B$-range to be Fourier transformed was chosen to include 10 full RSA periods as indicated by the yellow area in Fig. \ref{figSupp1}~(a), starting from and ending at the peaks of the outmost resonances. This yields a resolution of the FFT spectra of 10 points per period, as can be seen in Fig. \ref{figSupp1}~(b) where spectra for different FFT conditions are compared (see below). The $x$-axis is transformed from the  frequency $\omega = g \mu_B t / \hbar$ to a time axis $t$ in order to allow for an easier physical interpretation: this way each FFT spectrum represents the temporal evolution of the spin polarization at the respective pump-probe separation $\Delta x$.

The real parts of the FFT spectra of the simulated RSA traces are shown by the blue squares. Resonances occur at times $t_n = n T_{\mathrm{rep}}$ (corresponding to $\omega_n = n \omega_B = n g \mu_B T_{\mathrm{rep}} / \hbar$), as expected. For the ideal case $\Delta t = 0$ (upper spectra), all data points adjacent to the resonance peaks in the real part FFT spectrum are zero, such that the peak values are equal to the integrated resonance values (i.e. the sum over 5 data points for each resonance) shown by the open blue circles. For finite pump-probe delays $\Delta t < 0$, however, the data points beside the peaks in the real part FFT spectra deviate from zero, resulting in a decrease of the integrated FFT resonances (middle and lower spectrum). This decrease is strongest at small $t_n$, suggesting an apparently slower decrease of the spin amplitude with time and consequently an apparently larger spin dephasing time $T_2^*$. Note that this is an artefact of the FFT of RSA traces with finite pump-probe delay $\Delta t < 0$, and is not a physical effect as all RSA traces are simulated with the same input spin dephasing time $T_2^* = 120$~ns. The opposite behaviour is found for the integrated resonance values of the FFT power spectra (green circles), which are increased for finite $\Delta t <0$ as compared to the ideal case $\Delta t = 0$. Finally, the FFT can also be performed after applying a Hann window to the RSA traces, weighting the data around $B = 0$ most strongly, as indicated by the broken grey line in Fig. \ref{figSupp1}~(a) (upper curve). The power spectrum of the FFT with Hann window is shown by the red squares in Fig. \ref{figSupp1}~(b). Each FFT resonance is broadened to 3 data points with peaks half as high as those of the real part FFT spectra. In this case, the correct contribution of each spin packet to the RSA signal is found by the integrated resonances (red circles), which are almost the same for all pump-probe delays $\Delta t \leq 0$.

\begin{figure}[tb]
	\includegraphics[scale=1.20]{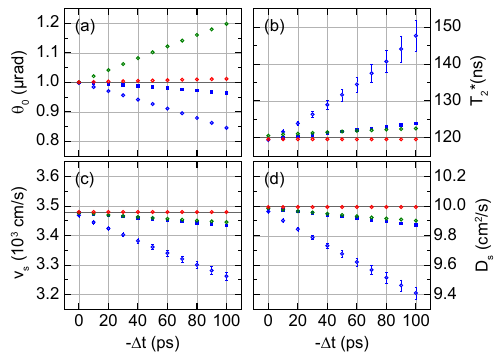}
	\caption{\label{figSupp2} Comparison of the spin transport parameteres resulting from the analysis model with different FFT conditions under the influence of finite pump-probe delays $\Delta t \leq 0$. Blue squares and open blue circles are the parameters resulting from the analysis of the peak values and of the integrated resonances of the real part FFT spectra, respectively, while the green and red circles result from the analysis of the integrated resonances of the FFT power spectra without and with Hann window, respectively. The parameters are (a) the spin amplitude $\theta_0$, (b) the spin dephasing time $T_2^*$, (c) the spind drift velocity $v_s$ and (d) the spin diffusion constant $D_s$. The input parameters of the simulation are represented by the dark grey lines.}
\end{figure}

In order to get a quantitative statement on the influence of the finite pump-probe delays $\Delta t <0$ on the spin transport parameters resulting from the different FFT methods, the lateral spin profiles are reconstructed from the FFT spectra as described in the main text, and $\theta_0$, $v_s$, $D_s$ and $T_2^*$ are extracted by Gaussian fits to these lateral spin profiles. The results are shown in Fig. \ref{figSupp2}. Here, the blue squares and open blue circles are the parameters resulting from the analysis of the peak values and of the integrated resonances of the real part FFT spectra, respectively, while the green and red circles result from the analysis of the integrated resonances of the FFT power spectra without and with Hann window, respectively. The input parameters of the simulation (identical for all FFT methods and for all pump-probe delays $\Delta t$) are represented by the dark grey lines. The analysis of the integrated resonances of the real part FFT spectra (open blue circles) shows the largest deviations of the extracted spin transport parameters from the input parameters. The best reproduction of the input parameters are found from the analysis of the integrated resonances of the FFT power spectra with Hann window (red circles), which shows almost no deviation from the input parameters for all the finite pump-probe delays.

\subsection*{3. Spin transport simulations for magnetic field-dependent spin dephasing times $T_2^*(B)$}
In a next step, spin transport simulations are performed with the same input parameters as before, see eq. (\ref{eq:InputParameters}), except that the pump-probe delay is fixed to $\Delta t = -50$~ps, as it was used for all measurements in the main text. Additionally, we use the following magnetic field dependence of the spin dephasing time $T_2^*(B)$:
\begin{eqnarray}
	\frac{1}{T_2^*(\Delta x, B)} &=& \frac{1}{\tau_0} + \frac{1}{\tau_B} \cdot \frac{\left|B\right|}{\mathcal{B} + \left|B\right|} \cdot \exp{\left[-\frac{\Delta x^2}{4\sigma^2}\right]}
	\label{eq:SimTauB01}\\
	\Rightarrow  T_2^*(0,B) &=& \frac{\tau_0}{1+\frac{\tau_0}{\tau_B} \frac{\left|B\right|}{\mathcal{B} + \left|B\right|}} =
	\begin{cases}
		\tau_0 & \textnormal{for} \ B = 0 \\
		\frac{\tau_0}{1+\frac{\tau_0}{\tau_B}} = \gamma \cdot \tau_0 \equiv \tau_{\infty} & \textnormal{for} \ \left|B\right| \gg \mathcal{B}.
	\end{cases}
\label{eq:SimTauB02}
\end{eqnarray}
This phenomenological form of the magnetic field-dependence of $T_2^*$ reflects the main features of the spin dephasing observed in the RSA scans in the main text. First, spin dephasing increases with the applied magnetic field strength $\left|B\right|$ but saturates beyond a saturation field $\mathcal{B}$, as can bee seen by the decrease of the RSA resonance height with $\left|B\right|$. Therefore, eq. (\ref{eq:SimTauB02}) describes a decrease of $T_2^*$ from the zero-field value $\tau_0$ at $B = 0$ to the saturation value $\tau_{\infty} = \gamma \tau_0$ at $\left|B\right| \gg \mathcal{B}$ (corresponding to $1/T_2^* = 1/\tau_0 + 1/\tau_B$), with $0 \leq \gamma \leq 1$. Note that the observed beaviour of the spin dephasing time cannot simply be attributed to inhomogeneous spin dephasing with a magnetic field-dependence $T_2^{*,\mathrm{(inh)}} \propto \left|B\right|^{-1}$, which does not saturate but continuously decreases to $\tau_{\infty}^{\mathrm{(inh)}} = 0$ for $\left|B\right| \rightarrow \infty$. To the best of our knowledge, a theoretical model describing this observed saturation of the magnetic field-dependence of the low-temperature spin dephasing time in n-GaAs at doping concentrations close the MIT does not exist. Second, the magnetic field-dependent decrease of $T_2^*(B)$ is strongest at the spatial pump-probe overlap $\Delta x = 0$ and becomes less pronounced at larger pump-probe separations $\left|\Delta x\right|$, as seen in the RSA-scans for different $\Delta x$ in the main text. This spatial dependence $T_2^*(\Delta x, B)$ is accounted for by the Gaussian function in eq. (\ref{eq:SimTauB01}). Possibly, this spin dephasing mechanism might originate from spin-spin-interactions of electrons which should scale with the spin density.

\begin{figure}[tb]
 \includegraphics{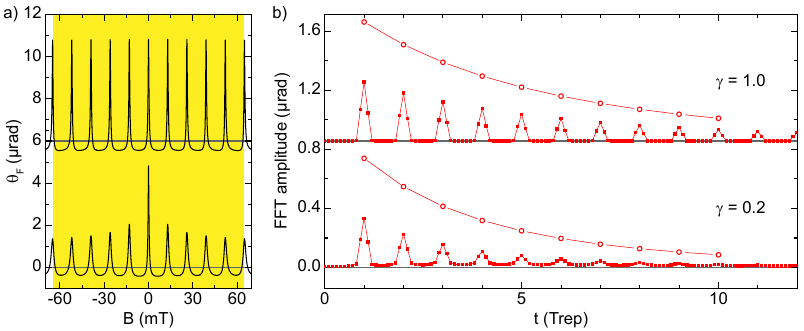}%
 \caption{\label{figSupp3} (a) Simulated RSA traces at $\Delta x = 0$ and $\Delta t = -50$~ps without and with strong magnetic field-dependent spin dephasing according to eqs. (\ref{eq:SimTauB01}) and (\ref{eq:SimTauB02}) using $\gamma = 1.0$ (upper curve) and $\gamma = 0.2$ (lower curve), respectively. (b) Corresponding FFT power spectra with Hann window (red squares) and integrated resonances (open red circles).}
\end{figure}

The phenomenological magnetic field-dependence $T_2^*(\Delta x,B)$ in eqs. (\ref{eq:SimTauB01}) and (\ref{eq:SimTauB02}) is characterized by the zero-field value, which is chosen to be $\tau_0 = 120$~ns as for the previous simulations, and by the saturation value $\tau_{\infty} = \gamma \tau_0$. Figure \ref{figSupp3}~(a) shows simulated RSA traces at $\Delta x = 0$ for $\gamma = \tau_{\infty}/\tau_0 = 1.0$, corresponding to the ideal case of no magnetic field-dependent spin dephasing with $\tau_{\infty} = \tau_0$ (upper curve), and $\gamma = 0.2$, corresponding to strong magnetic field-dependent spin dephasing (lower curve). In the latter case, the decrease and the broadening of the RSA resonances with $\left|B\right|$ is clearly visible. The FFT spectra for these simulated RSA traces are shown in Fig. \ref{figSupp3}~(b) with the red squares being the  FFT power spectra with a Hann window and the red circles the corresponding integrated resonances. Here, we restrict ourselves to the FFT method with Hann window because this method gave the best reproduction of the input parameters for finite pump-probe delays $\Delta t <0$, as discussed in the previous section. The influence of the magnetic field-dependent spin dephasing time $T_2^*(\Delta x, B)$ on the FFT spectra is seen as a broadening of the FFT resonances, compared to the ideal case $\gamma = 1.0$, as the data points in between resonances become non-zero. This is a typical behaviour for any form of damping. For example, let us consider the case of inhomogeneous dephasing, $T_2^{*,\mathrm{(inh)}} = \sqrt{2} \hbar /(\Delta g \mu_B \left|B\right|)$, due to a spread of $g$-factors $\Delta g$ around the average value $g_0$. A single spin packet then evolves as

\begin{equation}
	s_z(t) = s_0 \cdot \exp{\left[-\frac{t}{T_2^{*,\mathrm{(inh)}}}\right]} \cdot \cos{\left(g_0 \frac{\mu_B}{\hbar} B t\right)} \equiv s_0 \cdot \exp{\left[-\frac{\left|B\right|}{B_{\mathrm{inh}}}\right]} \cdot \cos{\left(g_0 \frac{\mu_B}{\hbar} B t\right)},
\end{equation}
such that the magnetic field-dependent spin dephasing acts like a damping of the spin precession, both in time and in the magnetic field. The Fourier transformation of such an exponentially damped oscillation yields Lorentzian resonances with finite width $\Delta \omega = B_{\mathrm{inh}}^{-1} = \Delta g \mu_B t / (\sqrt{2} \hbar)$. The FFT resonances in Fig. \ref{figSupp3}~(b) do not have Lorentzian shape as the phenomenological form of $T_2^*(\Delta x, B)$ in eq. (\ref{eq:SimTauB01}) does not lead to a simple, exponential damping of the spin precession in the magnetic field, but the broadening of the resonances can nevertheless be seen. It is accompanied by a faster decrease of the spin amplitude compared to the ideal case $\gamma = 1.0$, resulting in an apparently smaller spin dephasing time. This is expected, as the FFT analysis model includes a constant (i.e. magnetic field-independent) spin dephasing time, which will be the mean value $\left<T_2^*(B)\right>$ averaged over the considered $B$-range and taking into account the weighting of the RSA trace by the Hann window.

\begin{figure}[tb]
 \includegraphics[scale=1.20]{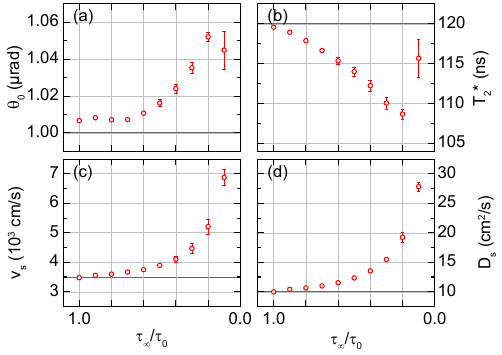}%
 \caption{\label{figSupp4} Influence of the magnetic field-dependent spin dephasing time on the resulting spin transport parameters (a) $\theta_0$, (b) $T_2^*$, (c) $v_s$ and (d) $D_s$. For a realistic value $\gamma = 0.8$ of the strength of the magnetic field-dependent spin dephasing, the deviations from the input parameters are less than 7~\%, showing the robustness of our FFT analysis model.}
\end{figure}

The results for the spin transport parameters from the application of our analysis model to the lateral spin profile which we reconstruct from the integrated resonances of the FFT power spectra with Hann window, are shown in Fig. \ref{figSupp4} as a function of $\gamma = \tau_{\infty}/\tau_0$, which represents the strength of magnetic field-dependent spin dephasing. Note that for smaller values of $\gamma$ the  field-dependent spin dephasing becomes stronger. $\theta_0$, $v_s$ and $D_s$ increase only slightly up to $\gamma \approx 0.5$ while more rapidly for $\gamma < 0.5$. In contrast, the extracted spin dephasing time $T_2^*$, decreases with increasing strength of the magnetic field-dependent spin dephasing.

The best agreement between the measured RSA traces in the main text and the simulated RSA traces presented here is found for $\gamma = 0.8$. For this value, $v_s$ is overestimated by 4~\% and $D_s$ by 7~\%, while $T_2^*$ is underestimated by less than 2~\%. This shows how robust our FFT analysis method is, even under systematic deviations from the ideal conditions assumed in the model.

\subsection*{4. Conclusions from the spin transport simulations}
As discussed above, the analysis of the lateral spin profiles reconstructed from the integrated resonances of the FFT power spectra with application of a Hann window give the most realistic spin transport parameters under real measurement conditions, i.e. for a finite pump-probe delay $\Delta t = -50$~ps and for a magnetic field-dependent spin dephasing time $T_2^*(\Delta x, B)$ according to eqs. (\ref{eq:SimTauB01}) and (\ref{eq:SimTauB02}) with $\gamma = 0.8$. These deviations from the assumptions made in the analysis model result in a systematic overestimation of $v_s$ by 4~\%, of $D_s$ by 7~\% and in a systematic underestimation of $T_2^*$ by 2~\%.

\end{document}